\documentclass[iop]{emulateapj}

\usepackage{color}
\usepackage[section]{placeins}
\usepackage{graphicx}
\usepackage{epstopdf}
\begin{document}

\title{
Model-independent constraints on cosmic curvature: implication from
updated Hubble diagram of high-redshift standard candles}

\author{Yuting Liu\altaffilmark{1}, Shuo Cao\altaffilmark{1$\ast$}, Tonghua Liu\altaffilmark{1}, Xiaolei Li\altaffilmark{2}, Shuaibo Geng\altaffilmark{1}, Yujie Lian\altaffilmark{1}, Wuzheng Guo\altaffilmark{1}}

\affiliation{1. Department of Astronomy, Beijing Normal University,
Beijing 100875, China; \emph{caoshuo@bnu.edu.cn} \\ 2. Department of
Physics, Hebei Normal University, Shijiazhuang 050024, China}

\begin{abstract}

The cosmic curvature ($\Omega_k$) is a fundamental parameter for
cosmology. In this paper, we propose an improved model-independent
method to constrain the cosmic curvature, which is geometrically
related to the Hubble parameter $H(z)$ and luminosity distance
$D_L(z)$. Using the currently largest $H(z)$ sample from the
well-known cosmic chronometers, as well as the luminosity distance
$D_L(z)$ from the relation between the UV and X-ray luminosities of
1598 quasars and the newly-compiled Pantheon sample including 1048
SNe Ia, 31 independent measurements of the cosmic curvature
$\Omega_k(z)$ can be expected covering the redshift range of
$0.07<z<2$. Our estimation of $\Omega_k(z)$ is fully compatible with
flat Universe at the current level of observational precision.
Meanwhile, we find that, for the Hubble diagram of 1598 quasars as a
new type of standard candle, the spatial curvature is constrained to
be $\Omega_k=0.08\pm0.31$. For the latest Pantheon sample of SNe Ia
observations, we obtain $\Omega_k= -0.02\pm0.14$. Compared to other
approaches aiming for model-independent estimations of spatial
curvature, our analysis also achieves constraints with competitive
precision. More interestingly, it is suggested that the
reconstructed curvature $\Omega_k$ is negative in the high redshift
region, which is also consistent with the results from the
model-dependent constraints in the literature. Such findings are
confirmed by our reconstructed evolution of $\Omega_k(z)$, in the
framework of a model-independent method of Gaussian processes (GP)
without assuming a specific form.

\end{abstract}

\keywords{cosmological parameters -- cosmology: observations}

\section{Introduction}

The spatial curvature of the Universe, i.e., whether the space of
our Universe is open, flat, or closed is one of the most fundamental
issues in  particle physics and modern cosmology. Its value, or even
its sign, is closely related to the fundamental Copernican principle
assumption and the Friedmann--Lema\^{\i}tre--Robertson--Walker
(FLRW) metric \citep{Qi19a,Cao19a}, an exact solution of the
Einstein's equations obtained under the assumption of homogeneity
and isotropy of space. Meanwhile, a possible detection of a nonzero
curvature also bears important information of many important
problems such as the evolution of our early universe
\citep{Ichikawa2006,Clarkson2007,Gong2007,Virey2008,Cao2019}, as
well as the accelerating expansion of the late-time universe, which
is supported by the observations of Type Ia suernovae (SNe Ia)
\citep{Riess98,Perlmutter99} in combination with independent
estimates of cosmic microwave background (CMB) \citep{Ade16},
ultra-compact structure in intermediate-luminosity radio quasars
\citep{Cao17a,Cao17b}, and strongly gravitationally lensing systems
(SGL) \citep{Cao12a,Cao12b,Cao15,Ma19}. Let us note that, although a
spatially flat universe is favored at very high confidence level by
the current popular observations (especially the latest Planck-2016
results of CMB \citep{Ade16}), the previous measurements of cosmic
curvature are indirect and model-dependent, which strongly depend on
a specific model for dark energy (e.g., the cosmological constant
$\Lambda$) \textbf{\citep{Valentino2019}}. However, considering the
strong degeneracy between the spatial curvature and the dark energy
equation of state, model-independent estimation for the spatial
curvature from different popular probes has been performed in the
literature \citep{Qi2019}.

The most straightforward technique to constrain the cosmic curvature
is by confronting the theoretical Hubble diagram (reconstructed by
the Hubble parameter measurements) with the observed luminosity
distances to the objects whose redshifts are known
\citep{Clarkson2008}. This test has been fully implemented with
updated observational SNe Ia data acing as standard candles
\citep{Shafieloo2010,Mortsell2011,Sapone2014,Cai2016}. However,
considering the uncertainty caused by nuisance parameters
characterizing SNe Ia light curves
\citep{Li2016c,Wei2017,Wang2017,Rana2017}, an improved
model-independent test of cosmic curvature to $z\sim 3.0$ has
recently performed with ultra-compact structures in radio quasars as
standard rulers \citep{Cao2019}. Meanwhile, \citet{Takada2015}
proposed that the combined radial and angular diameter distances
from the BAO can be used to constrain the curvature parameter, with
the achievable accuracy of such $\Omega_k$ measurement at $\Delta
\Omega_k\sim 10^{-3}$. Another method was also put forward to carry
out test in the framework of the sum rule of distances along null
geodesics of the FLRW metric, by employing strong lensing
observations (Einstein radius or time delays) and supernova distance
measurements \citep{Rasanen2015,Denissenya2018}. More recently, such
method has been applied to a latest data set of strong lensing
systems in combination with intermediate-luminosity quasars
calibrated as standard rulers \citep{Qi2019}. It is interesting to
note that, considering the cross-correlation between foreground mass
and gravitational shear of background galaxies, the assumed lens
model has a considerable impact on the cosmic curvature constraint,
which slightly favors a spatial closed Universe. In addition, some
recent studies have also discussed the possibility of extending the
above analysis to the simulated data of gravitational waves from
future gravitational wave detectors, which can be considered as
standard siren to provide the information of luminosity distance
\citep{Jimenez2018}.

In this paper, we will focus on a method that actually delivers
estimations of the curvature parameter at different redshift
\citep{Clarkson2007}, using the current observations of standard
candle data (quasars, SNe Ia) and standard clock data (Hubble
parameters $H(z)$ inferred from cosmic chronometers).We firstly
reconstruct the function of luminosity distance with respect to
redshift $z$ from two different standard candle data, depending on
the parameters characterizing the non-linear relation between the
X-ray and UV luminosities of quasars, as well as the light-curve
fitting parameters from SNe Ia sample. Next, with the Hubble
parameter measurements taken into consideration, we directly
transform the above observations to $\Omega_k$ at different
redshift, and thus achieve cosmological model-independent
constraints on the spatial curvature. Compared with the previous
works, measurements of $\Omega_k$ from observations at different
redshift could not only achieve a stringent measurement of the
spatial curvature in a direct geometric way, but also call into
doubt the FLRW metric and cosmic homogeneity and isotropy
\citep{Denissenya2018,Qi19a,Cao19a}. It is clear that, for the
purpose of implementing the method of \citet{Clarkson2007}, it would
be beneficial to use distance probes covering higher redshifts thus
taking advantage of larger sample of $H(z)$ observations. For $H(z)$
data, it can be derived from differential ages of galaxies [``cosmic
chronometer (CC)"] and the radial baryon acoustic oscillation (BAO)
scale in the galaxy distribution \footnote{Note that the expansion
rate measures obtained from BAO observations are possibly dependent
on the assumed fiducial cosmological model.}. In this analysis, we
update the largest distance data through the Hubble diagram of 1598
quasars ($z<5.100$) \citep{Risaliti2018} and the Pantheon catalog of
1048 SNe Ia ($z<2.3$) \citep{Scolnic2018}, based on which the cosmic
curvature at each specific redshift (corresponding to the redshift
of each Hubble parameter measurement) could be directly obtained.
When the latest quasar sample is used, the spatial curvature is
constrained to be $\Omega_k=0.08\pm0.31$. For the Pantheon SNe Ia,
we obtain $\Omega_k=-0.02\pm0.14$. These results, in the context of
model-independent estimations for spatial curvature, consistently
favor a spatially flat Universe. Finally, we use the
model-independent method Gaussian processes (GP) to reconstruct the
evolution of the curvature of the universe. Our results indicate
that a better quality of the observational data sets are also
required to detect a tiny cosmic curvature more precisely, which
will be also discussed in this paper.

This paper is organized as follows. In Sec. 2, we give a brief
introduction of the theoretical method and the data used in this
work. Sec. 3 investigates the constraints these data put on the
cosmic curvature. Finally, the conclusions are presented in Sect.~4.

\begin{figure}
\begin{center}
\includegraphics[width=1.0\linewidth]{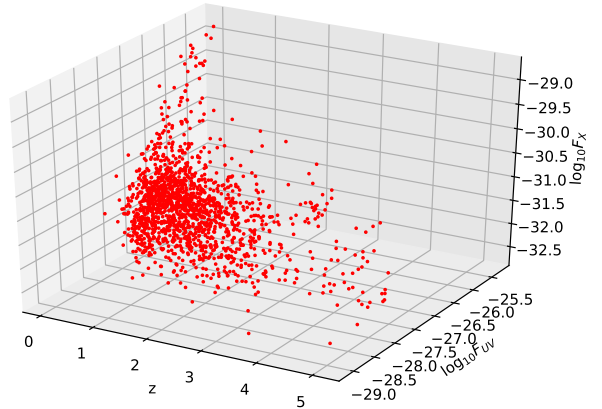}
\end{center}
\caption{The scatter plot of the X-ray and UV fluxes for 1598
quasars.}
\end{figure}

\begin{figure}
\begin{center}
\includegraphics[width=1.0\linewidth]{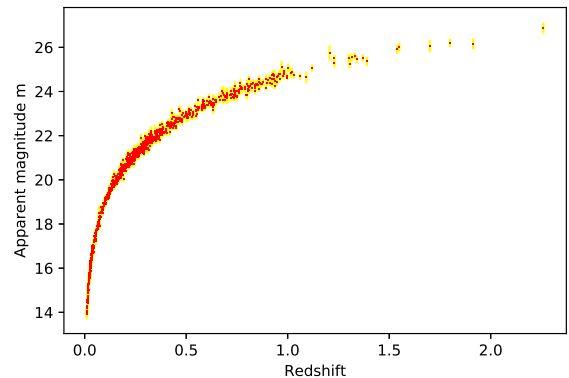}
\end{center}
\caption{The scatter plot of the 1048 SNe Ia Pantheon sample. The
red point denotes the apparent magnitude for each SNe Ia, with its
1-$\sigma$ confidence level (yellow bar).}
\end{figure}

\section{Method and observation data}

It is well known that in the FLRW metric, the luminosity distance
$D_L(z)$ can be expressed as
\begin{eqnarray}\label{dcadd}
D_L(z) &=& \frac{c(1+z)}{H_0\sqrt{|\Omega_{\rm k}|}}{\rm
sinn}\left[\sqrt{|\Omega_{\rm
k}|}\int_{0}^{z}\frac{dz'}{E(z')}\right]~,
\end{eqnarray}
where $H_0$ denotes the Hubble constant \footnote{In this work, we
adopt the prior of the Hubble constant $H_0=67.36\pm0.54$ km/s/Mpc
from the latest Planck CMB observations \citep{Ade16}.}, $c$ is the
speed of light, and $E(z)=H(z)/H_0$ is the dimensionless Hubble
parameter. The curvature parameter $\Omega_k$ is related to the
dimensionless curvature $K$ as $\Omega_k=-K c^2 /a_0^2 H_0^2$, where
$a_0$ is the present value of the scale factor, and $K=+1, -1, 0$
corresponds to closed, open, and flat universe. For convenience, we
denote ${\rm sinn}(x)=\sin(x),\,x,\,\sinh(x)$ for $\Omega_{\rm
k}<0,\,=0,\,>0$, respectively. The derivative of Eq.~(1) will
generate the cosmic curvature $\Omega_k$, which can be directly
determined by using the Hubble parameter and transverse comoving
distance as
 \citep{Clarkson2007}
\begin{equation}\label{eq1}
\Omega_k=\frac{[H(z)D'(z)]^2-c^2}{[H_0D(z)]^2},
\end{equation}
where $H(z)$ is the expansion rate at redshift $z$. The luminosity
distance $D_L(z)$ is simply related to the transverse comoving
distance $D(z)$ as $ D(z)=D_L(z)/(1+z)$ \citep{Hogg1999}, while
$D'(z)=dD(z)/dz$ denotes the derivative with respect to redshift
$z$. Thus we should use current observational data sets to
reconstruct $D(z)$ and $D'(z)$, independently, and combine these two
reconstructions with independent $H(z)$ measurements to derive
$\Omega_k$.

\subsection{Distance from quasars and type Ia supernovae observations}

As the brightest sources in the universe that can be observed up to
redshift $z\sim 8.0$, quasars \citep{Mortlock2011,Baados2018} have
long been considered as potential candidates for extending the
distance far beyond the limits imposed by supernovae ($z\sim 2.0$).
Different form the fundamental property of the cosmological standard
candle (standardized luminosity), the non-linear relation between
the X-ray and UV luminosities of quasars can be parameterized as a
linear dependence
\begin{equation}
\log_{10}(L_X)=\gamma \log_{10}(L_{UV})+\beta,
\end{equation}
where $L_X$ and $L_{UV}$ are the rest-frame monochromatic
luminosities at 2keV and 2500${\AA}$, while $\gamma$ and $\beta$
denote the slope parameter and the intercept. Based on the well
known flux -- luminosity relation, the non-linear relation between
the X-ray and UV fluxes of quasars can be written as
\begin{equation}
\log_{10}(F_X)=\gamma\log_{10}(F_{UV})+2(\gamma-1)\log_{10}(D_L)+\beta^{'},
\end{equation}
where $D_L$ is luminosity distance, $F_X$ and $F_{UV}$ represent the
X-ray and UV flux, and the intercept is rewritten as
$\beta^{'}=(\gamma-1)\log_{10}(4\pi)+\beta$. One may clearly see
that such relation could provide a  potential cosmological probe,
i.e., if there is no redshift evolution of the relation, the
observed X-ray flux is a function of the observed UV flux, the
redshift, and the luminosity distance. Therefore, relevant
cosmological parameters can be inferred by fitting this relation to
different quasar samples with multiple observations available
\citep{Risaliti2015,Lusso2016,Risaliti2017,Lusso2017}.

In this paper, we use the newly-built Hubble diagram of quasars from
a parent sample of 7237 sources, covering the redshift range of
$0.036<z<5.100$ \citep{Risaliti2018}. Note that three ``Cleaning''
criteria are used to derive the final 1598 sources from the parent
sample, in the framework of three filters including X-ray
absorption, observational contaminants in the UV, and Eddington
bias. The final sample is built merging five groups of quasars: 791
sources from the SDSS-DR7 sample, 612 sources from the SDSS-DR12,
102 sources from XMM-COSMOS, 18 sources from the low-redshift sample
(Swift), 19 sources from Chandra-Champ, 38 sources from the high-$z$
sample ($z>4$), and 18 sources from the new $z\sim 3$ sample
(XMM-Newton Very Large Program). The final results indicated that
such refined quasar sample could effectively mitigate the large
dispersion in the $L_X - L_{UV}$ relation, with more accurate slope
determination $\gamma=0.633\pm0.002$ and smaller dispersion
$\delta=0.24$. With a tractable amount of scatter avoiding possible
contaminants and unknown systematics, a Hubble diagram of quasars
could provide new measurements of the cosmic expansion at higher
redshifts ($z<5.10$), which has never been explored by any other
cosmological probes. The scatter plot of 1598 quasars is shown in
Fig~.1, using the most recent QSO compilation from
\citet{Risaliti2018}. Note that in this analysis, we will treat the
slope $\gamma$ and the intercept $\beta$ as two nuisance parameters.
The intrinsic dispersion is also taken as a free parameter
$\sigma_{int}$ contributing to the intrinsic scatter.

To reconstruct $D(z)$, we use SNe Ia Pantheon data set released by
the Pan-STARRS1 (PS1) Medium Deep Survey, which contains 1048 SNe Ia
data ranging from $0.01<z<2.3$ \citep{Scolnic2018}. Compared with
the previous SNe Ia samples extensively discussed in the previous
works, i.e., high-$z$ data ($z
>1.0$) from the SCP survey \citep{Suzuki2012}, the GOODS
\citep{Riess2007} and CAN-DELS/CLASH surveys
\citep{Rodney2014,Graur2014,Riess2017}, the Pantheon catalogue
extends the Hubble diagram to $z=2.26$, with the combination of the
subset of 279 PS1 SNe Ia \citep{Rest2014,Scolnic2014a} ($0.03 < z <
0.68$) and useful distance estimates of SNe Ia from SDS, SNLS,
various low-$z$ and HST samples \citep{Scolnic2018}. The observed
distance modulus of each SNe is given by
\begin{equation}
\mu=m_B - M_B + \alpha \cdot X _ { 1 }  -\beta^{*} \cdot \mathcal{C}
+ \Delta {M} + \Delta {B},
\end{equation}
where $m_B$ is the apparent \textit{B}-band magnitude, $M_B$ is the
absolute \textit{B}-band magnitude, $\mathcal{C}$ is the color
parameter quantifying the relation between luminosity and color, and
$X_{1}$ is the light-curve shape parameter quantifying the relation
between luminosity and stretch. Note that distance corrections based
on the host-galaxy mass ($\Delta {M}$) and predicted biases from
simulations ($\Delta {B}$) are also taken into account. Based on the
new approach called BEAMS with Bias Corrections (BBC)
\citep{Kessler2017}, the nuisance parameters in the Tripp formula
\citep{Tripp1998} were retrieved and the observed distance modulus
is simply reduced to $\mu=m_B - M_B$. We transform the distance
modulus $m_B-M_B$ given in the data set to $D_L$ using
\begin{equation}
D_{L}(z)=10^{\mu(z)/5-5} (\textmd{Mpc})
\end{equation}
Following the strategy of \citet{Scolnic2018}, six sources of
uncertainties are included in the distance modulus in Pantheon
dataset, i.e., the uncertainty from the photometric error
($\sigma_{\rm N}$), the mass step correction ($\sigma_{\rm Mass}$),
the distance bias correction ($\sigma_{\rm Bias}$), the peculiar
velocity uncertainty and redshift measurement uncertainty
($\sigma_{\mu - z}$), the lensing uncertainty due to the LOS mass
distribution ($\sigma_{\rm lens}$), and the intrinsic scatter
($\sigma_{\rm int}$). For this analysis, the total statistical
uncertainty is modeled as $\sigma^2_{SN}=\sigma^2_{\rm
N}+\sigma^2_{\rm Mass}+\sigma^2_{\mu - z}+\sigma^2_{\rm
lens}+\sigma^2_{\rm int}$ \citep{Scolnic2018}. The Pantheon SNe Ia
sample and the above error strategy has been widely applied to place
stringent limits on cosmological parameters \citep{Qi18}, provide
accurate measurements of the speed of light \citep{Cao18} and the
cosmic opacity at higher redshifts \citep{Ma19b,Qi19c}. Moreover, in
Fig.~2, we illustrate the dependence of apparent \textit{B}-band
magnitude on reshifts, derived from 1048 SNe Ia data covering the
redshift range of $0.01<z<2.3$.

Obtaining these observational data points of $D_L(z)$, we can use
different methods to reconstruct $D(z)$ and its derivative $D'(z)$.
In order to achieve model-independent estimation for the cosmic
curvature, we perform empirical fit to the  luminosity
distance measurements, based on a third-order logarithmic polynomial of
\citet{Risaliti2018},
\begin{equation}
D_L(z)=ln(10)c/H_0(x+a_1x^2+a_2x^3),
\end{equation}
where $x=\log(1+z)$, $a_1$ and $a_2$ are the two constants that need
to be optimized and determined by flux measurements of quasar data
and apparent \textit{B}-band magnitudes of SNe Ia data. Then, we
carry out the Markov Chain Monte Carlo (MCMC) method to obtain the
best-fit values and their uncertainties of parameters by using a
Python module called emcee \citep{Foreman-Mackey2013}
\footnote{https://pypi.python.org/pypi/emcee}. For the Hubble
diagram of quasar, the parameters ($\gamma$, $\beta$ and $\delta$
characterizing the $L_X - L_{UV}$ relation, $a_1$ and $a_2$
characterizing the luminosity distance) are optimized by minimizing
the $\chi^2$ objective function
\begin{equation}
\chi_{QSO}^2=\sum
\limits_{i=1}^{1598}{\frac{[\log_{10}(F_X)_i-\Phi([F_{UV}]_i,D_L[z_i])]^2}{\sigma_{QSO}^2}},
\end{equation}
where $\Phi([F_{UV}]_i,D_L[z_i])$ is defined as
\begin{equation}
\Phi([F_{UV}]_i,D_L[z_i])=\gamma\log_{10}([F_{UV}]_i)+2(\gamma-1)\log_{10}(D_L[z_i])+\beta^{'}
\end{equation}
The variance $\sigma_{QSO}^2=\delta^2 +\sigma_i^2$ is given in terms
of the global intrinsic dispersion ($\delta$), and the $i$-th
measurement error of $(F_X)_i$ \footnote{Note that the error in
$[F_{UV}]_i$ is presumed to be insignificant (compared with
$\sigma_i$ and $\delta$), which will be ignored in this analysis
\citep{Risaliti2015}.}. For the Hubble diagram of SNe Ia, the
parameters ($M_B$ characterizing the distance modulus, $a_1$ and
$a_2$ characterizing the luminosity distance) are optimized by
minimizing the $\chi^2$ objective function
 \begin{equation}
\chi_{SNe}^2=\sum \limits_{i=1}^{1048}{\frac{[m_i^{obs}-
m_i^{th}]^2}{\sigma_{SNe}^2}},
\end{equation}
where $\sigma_{SNe}$ accounts for error in SNe Ia observations
propagated from the covariance matrix \citep{Scolnic2018}. The
marginalized probability distribution of each parameter and the
marginalized 2-D confidence contours are presented in Figs.~3-4.

\begin{figure}
\begin{center}
\includegraphics[width=0.95\linewidth]{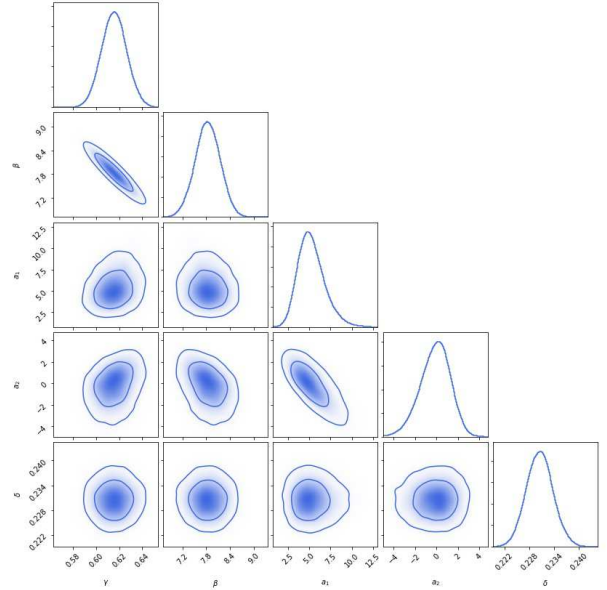}
\end{center}
\caption{Marginalized constraints on $a_1$, $a_2$, $\gamma$, $\beta$
and $\delta$ in the Hubble diagram of quasars.}
\end{figure}

\begin{figure}
\begin{center}
\includegraphics[width=0.95\linewidth]{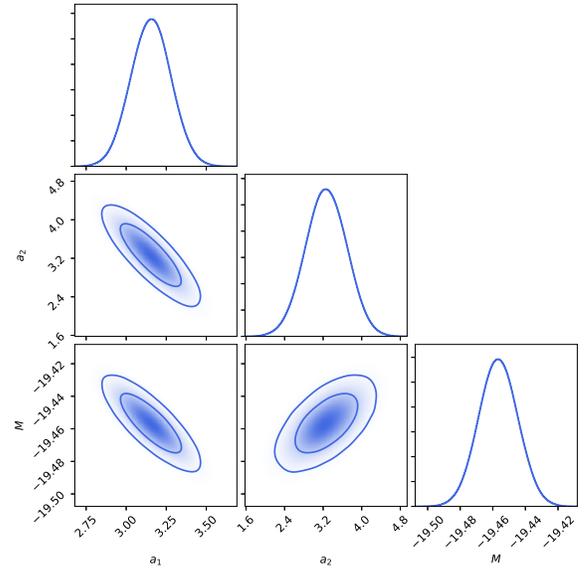}
\end{center}
\caption{Marginalized constraints on $a_1$, $a_2$ and $M$ in the
Pantheon sample.}
\end{figure}

\begin{figure}
\begin{center}
\includegraphics[width=0.95\linewidth]{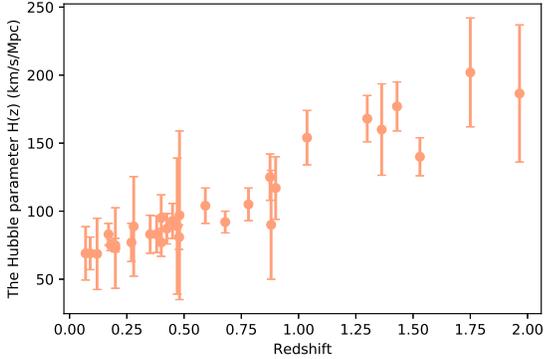}
\end{center}
\caption{The latest measurements of 31 Hubble parameters from the
galaxy differential age method.}
\end{figure}

\subsection{The expansion rate measurements $H(z)$ }

The expansion rate at any redshifts, i.e., the Hubble parameter is
defined as $H(z)$=$\dot{a}$/$a$, where $a$ denotes the scale factor
and $\dot{a}$ represents its derivative with respect to cosmic time
$t$. \citet{Jimenez2002} proposed a model-independent method to
calculate the expansion rate of the universe by the differential age
evolution of passively evolving galaxies
\begin{equation}
H(z)=\frac{\dot{a}}{a}=-\frac{1}{1+z}\frac{dz}{dt}.
\end{equation}
By measuring the age difference between the two galaxies at
different redshifts, the Hubble parameter $H(z)$ can be directly
obtained from the so-called differential age (DA) or cosmic
chronometer approach (in which one calculates the value of $dz/dt$).
Actually, $H(z)$ data can also be obtained through the detection of
radial baryon acoustic oscillations (BAO) from galaxy clustering in
redshift surveys
\citep{Gaztaaga2009,Blake2012,Busca2013,Samushia2013,Xu2013,Font-Ribera2014,Delubac2015}.
However, some recent studies indicated that the expansion rate
measurements obtained from BAO observations are possibly dependent on
the assumed fiducial cosmological model and the prior for the
distance to the last scattering surface from CMB observations
\citep{Li2016c}. Therefore, in our analysis, we use only the latest
31 DA $H(z)$ measurements in the redshift range $0.070<z<1.965$,
which is compiled and presented in Fig.~5.

Now we combine the reconstructions of $D(z)$ and $D'(z)$ with $H(z)$
measurements, which can be applied to the derivation of
$\Omega_k(z)$ in Eq.~(2). We stress again that the cosmic curvature
test is model independent, so we need not assume any cosmological
model, and the two data sets of cosmic chronometers and standard
candles are also independent of each other.

\subsection{The Gaussian Processes (GP)}

In order to reconstruct the evolution of the cosmic curvature
$\Omega_k(z)$, the Gaussian processes (GP) method will be adopted in
the following analysis, with which one can perform a reconstruction
of a function and its derivatives from a given data set without
assuming any cosmological models. Such approach, which was
originally proposed by \citet{Seikel2012a} and extensively applied
in various studies in the literature
\citep{Shafieloo12,Cao19a,Qi19a,Liu19,Wu2020,Zheng2020}, is
particularly useful to describe the observed data using the
distribution over functions provided by GP. On the assumption that
each data point satisfies a Gaussian distribution and the full
observational data set follows a multivariate normal distribution,
the value of a function $f(z)$ evaluated at a point $z$ depends on
the mean value $\mu(z)$ and the covariance function Cov$(f(z)$,
$f$($\tilde{z}$))=k($z$, $\tilde{z}$). In the framework of such
mathematical formalism, a random function $f(z)$ without any
observations can be generated using the covariance matrix from the
GP, based on the observational data and the values of the
corresponding slopes at $z$. For a set of input points $Z=\left\{
z_i \right\}$, one can generate a vector of function values at $Z^*$
as
\begin{equation}
f^*=\mathcal{N}(\mu^*,K(Z^*, Z^*)).
\end{equation}
Similarly, the observational data can be written as
\begin{equation}
Y=\mathcal{N}(\mu,K(Z, Z)+C),
\end{equation}
where $C$ is the covariance matrix of the data. Using the values of
$y$ at $Z$, one can reconstruct the mean and covariance of $f^*$ as
\begin{equation}
\overline{f^*}=\mu^*+K(Z^*, Z)[K(Z, Z)+C]^{-1}(y-\mu)
\end{equation}
and
\begin{equation}
{\text{Cov}}(f^*)=K(Z^*, Z^*)-K(Z^*, Z)[K(Z, Z)+C]^{-1}K(Z, Z^*),
\end{equation}
Note that the derivative of the function $f(z)$ can also be
calculated through the covariance function.

The crucial task in Gaussian process techniques is to determine the
covariance function, with which one can derive the quantities at
some redshifts at which they have not been directly measured. In
this paper, we focus on the squared exponential covariance function
to correlate the values of cosmic curvature at the two different
redshifts ($z$ and $\tilde{z}$):
\begin{equation}
k(z,\tilde{z})=\sigma^2_{f}exp(-\frac{(z-\tilde{z})^2}{2\ell^2}),
\end{equation}
Here $\ell$ quantifies the characteristic length in x-direction to
get a significant change in $f(z)$, whereas $\sigma_{f}$ denotes the
corresponding typical change in the y-direction. The two
hyperparameters ($\ell$ and $\sigma_{f}$) characterizing the
bumpiness of the function can be constrained from the observational
data. It should be pointed out that compared with other choices of
covariance functions (the Mat\'{e}rn and Cauchy covariance
function), the advantage of the squared exponential covariance
function lies in its effective reconstruction of the derivative of a
function \citep{Seikel2012a}. Such issue has been extensively
discussed in the recent studies of \citet{Zheng2020}, which also
found the insignificant differences between reconstructions
performed with different covariance functions (the Mat\'{e}rn,
Cauchy, and the squared exponential covariance function). Therefore,
in the following analysis zero mean function and squared exponential
covariance function will be applied to obtain the reconstructed
$\Omega_k(z)$. Such model-independent method is executed in the
publicly available code called GaPP (Gaussian Processes in Python)
\footnote{http://ascl.net/1303.027}.

\section{Results and discussion}

By applying the above mentioned procedure to the quasar distance
reconstruction and 31 Hubble parameter measurements, we obtain the
results shown in Fig.~6 to determine the $\Omega_k(z)$ at each point
$z$, which we want to reconstruct. The uncertainty of these
measurements are calculated from propagated uncertainties of $D(z)$,
$D'(z)$ and $H(z)$. In principle, the function $\Omega_k(z)$ can be
reconstructed from observations, and the FLRW metric is ruled out if
$\Omega_k(z)$ is not constant. However, at lower redshifts, the
errors become large due to the poor reconstructions of $D(z)$ and
$D'(z)$ in that region. The accuracy of these measurements improves
with increasing redshift $z$. In Fig.~6, it is shown that all of the
reconstructed $\Omega_k(z)$ is consistent with the vanishing cosmic
curvature within the 1$\sigma$ limit. Therefore, estimation of the
spatial curvature using $H(z)$ and quasars is fully compatible with
flat Universe at the current level of observational precision. Then
we can give the weighted mean of the present value of curvature
density parameter, based on the most straightforward and popular way
of summarizing multiple measurements, i.e., inverse variance
weighting \citep{Cao19a}
\begin{equation}
\begin{array}{l}
\Omega_k=\frac{\sum\left(\Omega_{k,i}/\sigma^2_{\Omega_{k,i}}\right)}{\sum1/\sigma^2_{\Omega_{k,i}}},\\
\sigma^2_{\Omega_{k}}=\frac{1}{\sum1/\sigma^2_{\Omega_{k,i}}},
\end{array}
\end{equation} where $\Omega_{k}$ stands for the weighted mean
of cosmic curvature and $\sigma_{\Omega_{k}}$ is its uncertainty. We
find that, from the quasar and cosmic chronometer observations,
model-independent estimation for the spatial curvature is
$\Omega_k=0.08\pm0.31$. This is fully in agreement with the
constraints obtained from the latest Planck CMB measurements
\citep{Ade16}. Moreover, one issue which should be discussed is the
comparison of our cosmological results with those of earlier studies
done using alternative probes. More specifically, the precision of
this estimation is comparable to that derived from the current
estimation of the cosmic curvature from the recently compiled set of
120 intermediate-luminosity quasars (ILQSO) observed in a
single-frequency VLBI survey \citep{Cao2019,Qi2019}. Such conclusion
is also well consistent with the recent analysis of
\citet{Rasanen2015}, which discussed constraints on cosmic curvature
by combining the strong lensing and supernova distance measurements,
in the framework of another model-independent test based on the
distance sum rule.

\begin{figure}
\begin{center}
\includegraphics[width=0.95\linewidth]{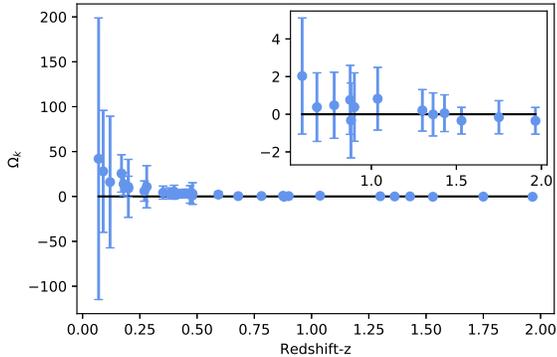}
\end{center}
\caption{31 measurements of the spatial curvature parameter
$\Omega_k$ and its details (upper right), from the Hubble diagram of quasars and expansion rate
measurements of cosmic chronometer.}
\end{figure}

\begin{figure}
\begin{center}
\includegraphics[width=0.95\linewidth]{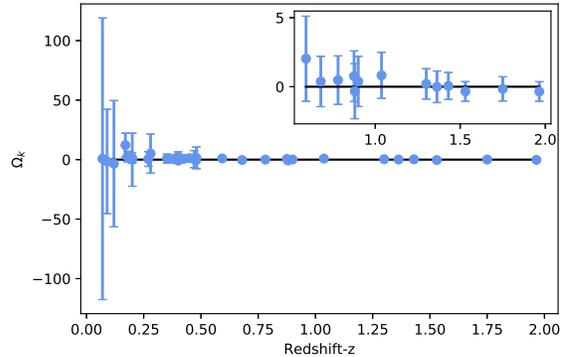}
\end{center}
\caption{31 measurements of the spatial curvature parameter
$\Omega_k$ and its details (upper right), from the Hubble diagram of  Pantheon SNe Ia sample and expansion rate
measurements of cosmic chronometer}
\end{figure}

When the Pantheon SNe Ia is used, it would increase chance of
finding significantly different $\Omega_{k}$ at different redshifts,
in the case when FLRW metric breaks down on some large scale. The
results are shown in Fig.~7. More importantly, we obtain that the
spatial curvature is model independently constrained to be
$\Omega_{k}=-0.02\pm0.14$, which suggests that there is no
significant signal to indicate the deviation of the cosmic curvature
$\Omega_{k}$ from zero at the current observational data level
[$H(z)$ and SNe Ia]. Compared with what obtained from the quasars,
there is an improvement in precision when the Pantheon SNe Ia is
considered, in the context of model-independent testing of the
cosmic curvature. However, there are several sources of systematics
we do not consider in the above analysis and which remain to be
clarified for this methodology. Specially, it is apparent that
sample incompleteness, which affects the number of available $H(z)$
measurements, will also play an important role in the $D(z)$+$D'(z)$
reconstruction and thus the effectiveness of this model-independent
test (A tiny change in the reconstructed  $D(z)$ and $D'(z)$ would
result in a very significant change to the nearby $\Omega_k$
measurements). In order to investigate the impact of sample
incompleteness on $\Omega_k$ estimation, we also carry out the
analysis by dividing the full sample into different sub-samples
given their redshifts and fitting a constant $\Omega_k$ in each
subsample. The redshifts of the $H(z)$ data span from $z=0.07$ to
$z=1.965$, so we divide the $H(z)$ measurements into five groups
with $z<0.5$, $0.5<z<1.0$, $1.0<z<1.5$, and $1.5<z<2.0$. The first
group has 19 $H(z)$ measurements with redshifts $z<0.5$, the second
group has 5 $H(z)$ with redshifts $0.5<z<1.0$, the third group has 4
$H(z)$ with redshifts $1.0<z<1.5$, and the fourth group contains 3
$H(z)$ with $1.5<z<2.0$. The cosmic curvature parameter can be
obtained as $\Omega_k=0.34\pm0.81$, $\Omega_k=0.03\pm0.38$,
$\Omega_k=0.31\pm0.26$, and $\Omega_k=-0.22\pm0.19$ at 68.3\%
confidence level, respectively. Note that the derived curvature is
negative in the high-redshift region, which is also consistent with
the results from the model-dependent constraints in the literature
\citep{Cai2016}.

Finally, an accurate reconstruction of $\Omega_k(z)$ can
considerably improve our understanding of the inflation models and
fundamental physics \citep{Cai2016}. In order to investigate the
evolution of $\Omega_k(z)$ without assuming a specific form, a
model-independent method of Gaussian processes (GP)
\citep{Seikel2012a} can be employed to reconstruct the cosmic
curvature from the observational data straightforwardly, without any
parametric assumption regarding the cosmological model.
 Fig.~8 shows the $\Omega_k$ parameter as a function
of redshift for the two different cases, derived from the combined
data sets of $H(z)$+QSO (upper) and $H(z)$+SNe Ia. One could note
that a universe with zero curvature (spatially flat geometry) is
strongly supported by the available observations. This is the most
unambiguous result of the current data sets. Moreover, the accuracy
of these measurements strongly depend on redshift $z$ and the
quality of the observational data, including the Hubble parameter
measurements, quasar and SNe Ia sample. We expect that as the
precision of the future data improves, especially at higher
redshifts, our approach will yield an even more accurate
determination of $\Omega_k$. Interestingly, although the constraint
by using quasars flux measurements data has not obvious improving
compared to using SNe Ia, it help us have a deeper understanding of
the cosmic curvature at earlier stage of the universe. Such issue
has been extensively discussed in many previous works
\citep{Cao2019,Qi2019}.

\section{Conclusions}

In this paper, we have used a new model-independent method to test
the cosmic curvature, based on the current $H(z)$ observations from
the well-known cosmic chronometers, as well as the luminosity
distance $D_L(z)$ from the relation between the UV and X-ray
luminosities of quasars \citep{Risaliti2018} and the newly-compiled
SNe Ia data (Pantheon sample) \citep{Scolnic2018}. Our results show
that 31 independent measurement of the cosmic curvature can be
expected covering the redshift range of $0.07<z<2$ and the approach
initiated in \citet{Clarkson2007} can be further developed. Firstly,
we reconstruct a function of luminosity distance $D_L(z)$ and its
derivative $D'(z)$, with the currently largest compilation of two
types of standard candles (quasars and SNe Ia). In order to dodge
the reliance on any assumptions of cosmological model, we directly
derive the Hubble diagram of standard candles from a third-order
logarithmic polynomial, with its undetermined coefficients
simultaneously fitted with other nuisance parameters (which
characterizes the $L_X - L_{UV}$ relation of quasars and the light
curve of SNe Ia). Then, one can calculate the curvature parameter
$\Omega_k$ using expansion rate $H(z)$ measurements obtained from a
sample of cosmic chronometers observations.

\begin{figure}
\begin{center}
\includegraphics[width=0.9\linewidth]{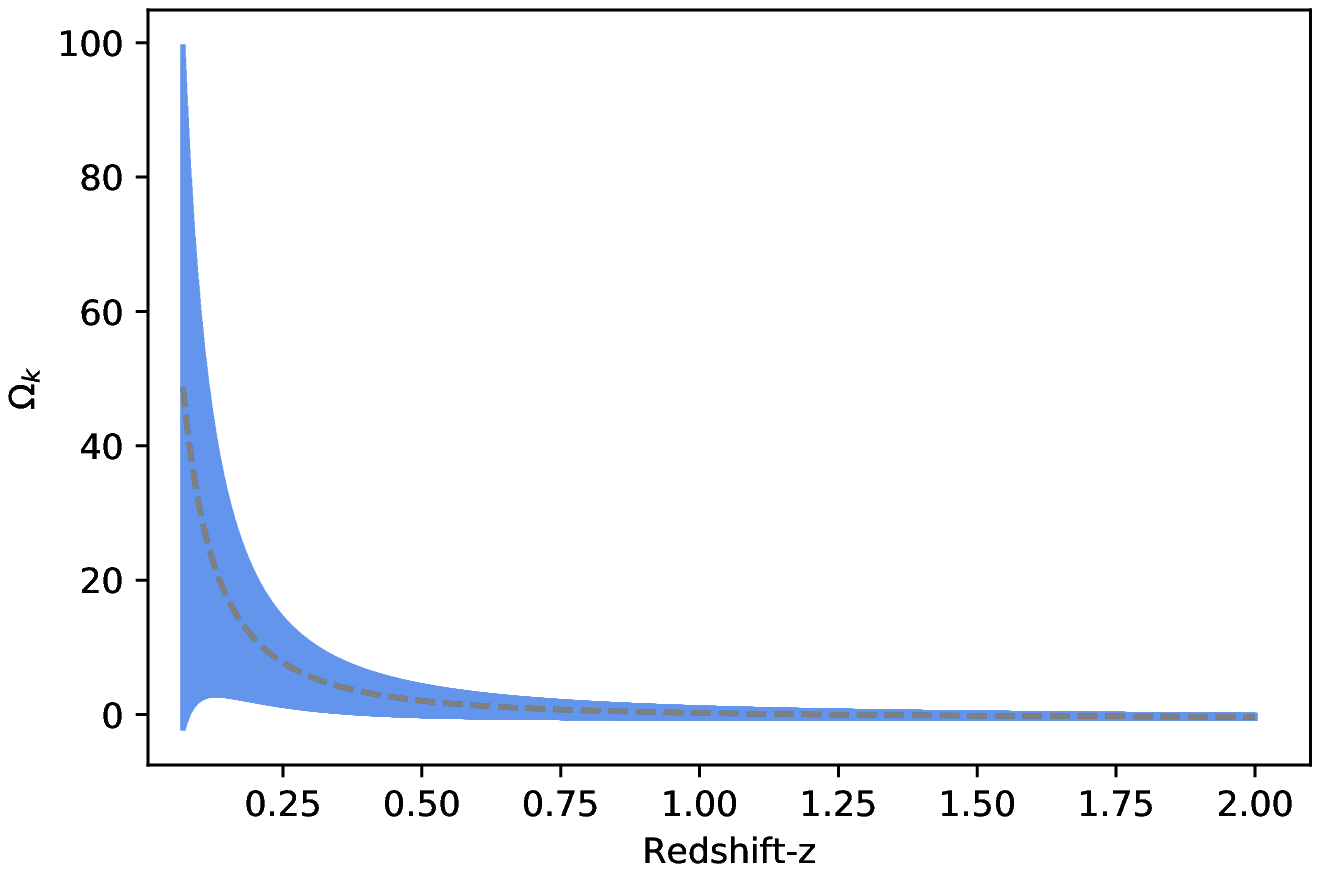}
\includegraphics[width=0.9\linewidth]{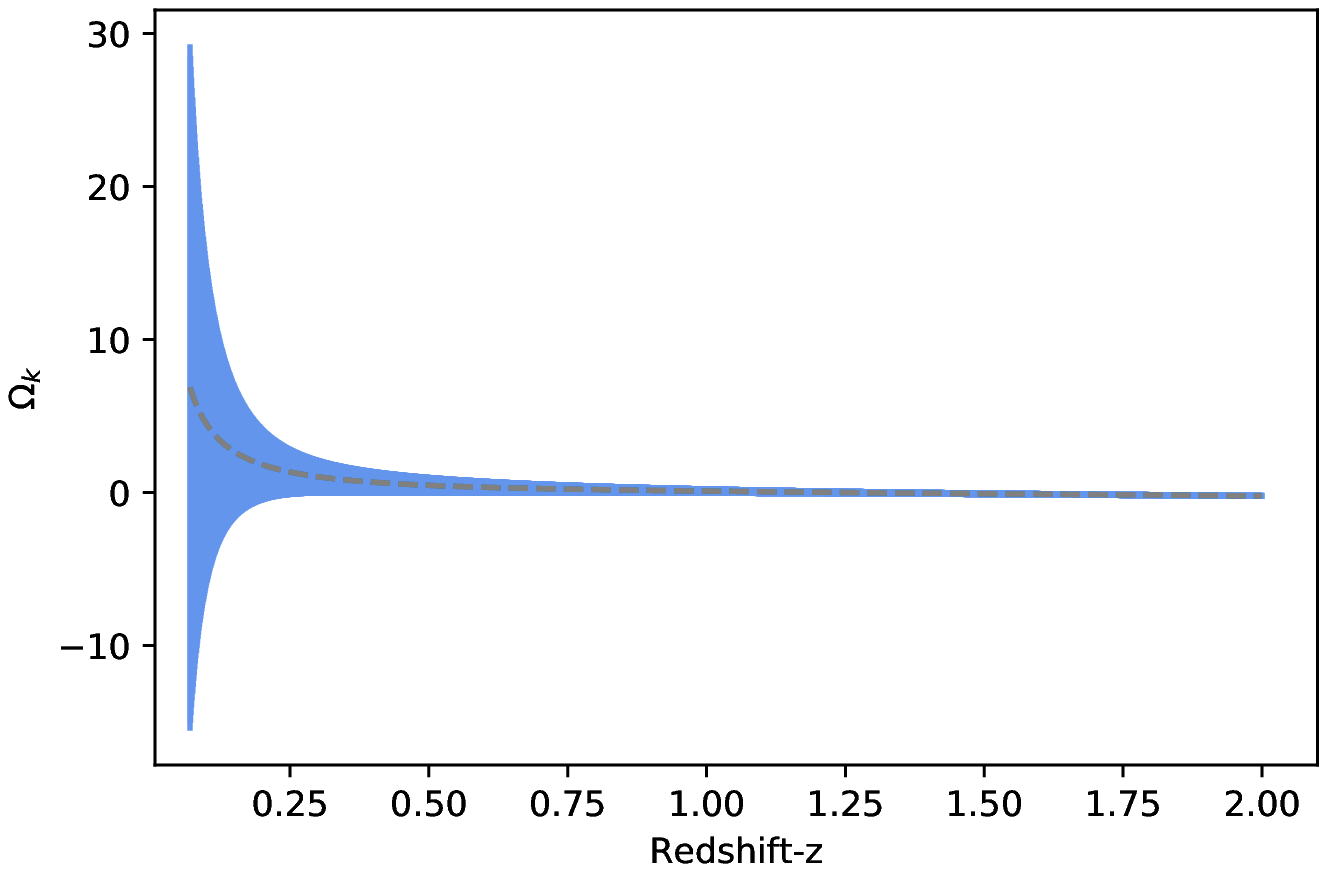}
\end{center}
\caption{Gaussian process reconstruction of $\Omega_k$ obtained from
the combined data sets of $H(z)$+QSO (upper) and $H(z)$+SNe Ia
(lower). The blue regions are the 68\% C.L. of the reconstructions.}
\end{figure}

By applying the above mentioned procedure to the QSO+$H(z)$ and SNe
Ia+$H(z)$ observations, we have shown that all of the calculated
$\Omega_k(z)$ is consistent with the vanishing cosmic curvature
within the 1$\sigma$ limit. Therefore, our estimation of the spatial
curvature is fully compatible with flat Universe at the current
level of observational precision. We find that the spatial curvature
is constrained to be $\Omega_k=0.08\pm0.31$ when the quasar and
cosmic chronometer observations are used. Such result in our
analysis is improved in precision, compared to the latest
model-independent estimations of the spatial curvature with the
distance sum rule \citep{Rasanen2015}. When the Pantheon SNe Ia is
considered, more stringent constraints on the cosmic curvature could
be achieved ($\Omega_k= -0.02\pm0.14$), although the mean of the
reconstructed curvature $\Omega_k$ is negative in the high redshift
region, which is also consistent with the results from the
model-dependent constraints in the literature \citep{Cai2016}. In
order to investigate the evolution of $\Omega_k(z)$ without assuming
a specific form, a model-independent method of GP
\citep{Seikel2012a} is employed to reconstruct the cosmic curvature
from the observational data straightforwardly. Our results show that
a universe with zero curvature (spatially flat geometry) is strongly
supported by the available observations, which is the most
unambiguous result of the current data sets.

Future observations will improve the constraints on the cosmic
curvature. On the one hand, properly calibrated UV - X-ray relation
in quasars has a great potential of becoming an important and
precise distance estimator in cosmology. A more accurately measured
quasar sample observed by SDSS \citep{Shen2011,Paris2017} and XMM
\citep{Rosen2016}, particularly at high redshifts, should provide an
even more stringent constraint on $\Omega_k$. On the other hand,
more accurate measurements of Hubble parameters may also improve the
effectiveness of our approach in the future. For instance, the
Extended Baryon Oscillation Spectroscopic Survey (eBOSS) will
compile 250,000 new, spectroscopically confirmed luminous red
galaxies, which yield measurements of $H(z)$ with 2.1\% precision
\citep{Dawson16}. These upcoming improvements on the precision of
cosmic curvature estimation will be of great significance for
understanding the evolution of the Universe and the nature of dark
energy.

\section*{Acknowledgments}

We would like to thank Jingzhao Qi and Xiaogang Zheng for their
helpful discussions.  This paper is dedicated to the 60th
anniversary of the Department of Astronomy, Beijing Normal
University. This work was supported by the National Key R\&D Program
of China No. 2017YFA0402600; the National Natural Science Foundation
of China under grant Nos. 11690023, 11633001 and 11920101003; the
Strategic Priority Research Program of the Chinese Academy of
Sciences, grant No. XDB23000000; Beijing Talents Fund of
Organization Department of Beijing Municipal Committee of the CPC;
the Interdiscipline Research Funds of Beijing Normal University; and
the Opening Project of Key Laboratory of Computational Astrophysics,
National Astronomical Observatories, Chinese Academy of Sciences. X.
Li is also supported by National Natural Science Foundation of China
under Grants No. 11947091, and the fund of Hebei Normal University
under Grants No. L2020B20.

\end{document}